\newcommand{\pl}{\partial}\newcommand{\ov}{\over}
 \newcommand{\noi}{\noindent}
\newcommand{\bq}{\begin{equation}} \newcommand{\eq}{\end{equation}}
\newcommand{\ra}{\rightarrow} 
\newcommand{\ph}{\phi} \newcommand{\iy}{\infty}
\newcommand{\hf}{{1\ov2}} \newcommand{\sk}{\sum_{i+j=k-1}(-1)^i\,}
\newcommand{\smk}{\sum_{i+j=-k-1}(-1)^{i+1}\,}
\newcommand{\um}{u_{-1}} \newcommand{\vm}{v_{-1}}

\renewcommand{\sp}{\vspace{2ex}}
\newcommand{\tn}{\otimes}

\documentstyle[12pt]{article} \textwidth=6in \oddsidemargin=.3in
\begin{document}
\begin{center} {\large\bf Fredholm Determinants} \end{center}
\begin{center}{\large\bf and the}\end{center}
\begin{center} {\large\bf mKdV/Sinh-Gordon Hierarchies} \end{center}
\sp\begin{center}{{\bf Craig A. Tracy}\\
{\it Department of Mathematics and Institute of Theoretical Dynamics\\
University of California, Davis, CA 95616, USA}}\end{center}
\begin{center}{{\bf Harold Widom}\\
{\it Department of Mathematics\\
University of California, Santa Cruz, CA 95064, USA}}\end{center}\sp
\begin{abstract}
For a particular class of integral operators $K$ we show that
the quantity \[\ph:=\log\,\det\,(I+K)-\log\,\det\,(I-K)\] satisfies both
the integrated mKdV hierarchy and the sinh-Gordon hierarchy. This proves
a conjecture of Zamolodchikov.
\end{abstract}
\sp
\noi{\bf I. Introduction}\sp

In recent years it has become apparent that there is a fundamental
connection between certain Fredholm determinants
and total systems of differential equations.
This connection first appeared in work on the scaling limit
of  the 2-point correlation function in the 2D Ising model \cite{mtw,wmtb}
and the subsequent generalization to  $n$-point correlations
and  holonomic quantum fields \cite{smj}.
In applications the Fredholm determinants
 are either correlation functions or closely related to correlations functions
in various statistical mechanical or quantum field-theoretic models.
In the simplest of
cases the differential equations are one of the Painlev{\'e} equations.
Some, but by no means a complete set of, references to these
further developments are
\cite{B&LC,its90,JMMS,KBI,tw5,tw5_5,Z}.  The review paper \cite{mccoy}
can be consulted for more examples of this connection.
\par
In recent work by the present authors on random matrices, techniques
were developed that gave simple proofs
of the connection between a large
class of Fredholm determinants and differential
equations \cite{tw5,tw5_5}.
In this paper we show how the philosophy of \cite{its90,KBI,tw5,tw5_5}
 can be applied to study
Fredholm determinants which are associated with operators $K$ having kernel
of the form
\[K(x,y)={E(x)\,E(y)\ov x+y}\]
where
\[E(x)=e(x)\,\exp\Big(\sum\hf\,t_k\,x^k\Big).\]
The (finite) sum is taken over odd positive and negative integers $k$. The
domain
of integration for the operator is $(0,\,\iy)$, and the function $e(x)$ can be
very
general. All that is required is that the operator be trace class for a range
of
values of the $t_k$ so the Fredholm determinants are defined. The quantity of
interest is
\bq\ph:=\log\,\det\,(I+K)-\log\,\det\,(I-K).\label{phdef}\eq
We shall show that $\ph$ satisfies the equations of the integrated mKdV
hierarchy
if $t_1$ is the space variable and $t_3,\,t_5,\cdots$ the time variables, and
that it satisfies the sinh-Gordon
hierarchy when $t_{-1},\,t_{-3},\cdots$ are the time variables.

To state the results precisely, the first assertion is that for $n\geq1$,
\bq {\pl \ph\ov\pl t_{2n+1}}=
(D^2-4\,{\pl\ph\ov\pl t_1}\,D^{-1}\,{\pl\ph\ov\pl t_1}
\,D)^n\,{\pl\ph\ov\pl t_1},\label{mKh}\eq
where $D$ denotes $\pl/\pl t_1$ and $D^{-1}$ denotes the antiderivative which
vanishes at $t_1=-\iy$. (Observe that $\ph$ and all its derivatives vanish
at $t_1=-\iy$.) This is the integrated mKdV hierarchy of equations,
\[{\pl\ph\ov \pl t_3}={\pl^3\ph\ov\pl t_1^3}-2\,({\pl\ph\ov\pl t_1})^3,\]
\[{\pl\ph\ov \pl t_5}={\pl^5\ph\ov\pl t_1^5}-10\,({\pl^2\ph\ov\pl t_1^2})^2\,
{\pl\ph\ov\pl t_1}-10\,({\pl\ph\ov\pl t_1})^2\,{\pl^3\ph\ov\pl t_1^3}+
6\,({\pl\ph\ov\pl t_1})^5,\]
etc. (In general there are constant factors on the left sides which can be
removed by changes of scale in the time variables; e.g.~\cite{AM})

To go in the other direction we introduce the inverse of the operator appearing
in (\ref{mKh}), which is given by
\bq(D^2-4\,{\pl\ph\ov\pl t_1}\,D^{-1}\,{\pl\ph\ov\pl t_1}\,D)^{-1}=
\hf(D^{-1}\,e^{2\ph}\,D^{-1}\,e^{-2\ph}+D^{-1}\,e^{-2\ph}\,D^{-1}\,e^{2\ph}).
\label{inv}\eq
(Precisely, this is the inverse in a suitable space of functions. See Lemma 4
below.)
We shall show that for $n\geq1$ we have the sinh-Gordon hierarchy of equations
\bq{\pl\ph\ov\pl
t_{-2n+1}}=2^{-n}\,(D^{-1}\,e^{2\ph}\,D^{-1}\,e^{-2\ph}+D^{-1}\,
e^{-2\ph}\,D^{-1}\,e^{2\ph})^n\,{\pl\ph\ov\pl t_1}.\label{sgh}\eq
The case $n=1$ of this is equivalent to the sinh-Gordon equation
\bq {\pl^2\ph\ov\pl t_{-1}\pl t_1}={1\ov2}\sinh\,2\ph.\label{sg}\eq
Observe that (\ref{mKh}) and (\ref{sgh}) can be combined into the single
statement
that either of them holds for all values of the integer $n$.  Further observe
that these results hold independently of the function $e(x)$ appearing in the
kernel $K(x,y)$.  The function $e(x)$ affects the boundary conditions for
(\ref{mKh})
and (\ref{sgh}) at $t_k=-\infty$.

That $\ph$ satisfies the integrated mKdV hierarchy was conjectured in \cite{Z},
and
that it satisfies the sinh-Gordon equation (\ref{sg}) was conjectured in
\cite{Z} and
proved in \cite{B&LC}.   A related identity,
\bq -{\pl^2\ov\pl t_{-1}\pl t_1}\log\det\,(I-K)={e^{2\ph}-1\ov4},\label{I-K}\eq
was also conjectured in  \cite{Z}  and proved in \cite{B&LC}, and will be
rederived
here.

We prove our results by expressing all relevant quantities in terms of inner
products
\bq u_{i,j}:=((I-K^2)^{-1}E_i,\,E_j),\qquad
v_{i,j}:=((I-K^2)^{-1}\,K\,E_i,\,E_j),
\label{uvdef}\eq
where $E_i(x):=x^i\,E(x)$, and showing that these quantities satisfy nice
differentiation
and recursion formulas.  Observe that both
$u_{i,j}$ and $v_{i,j}$ are symmetric in the
indicies since the operator $K$ is symmetric.
That these inner products are basic is expected from earlier investigations;
e.g.~\cite{its90,KBI,tw5,tw5_5}.\sp

\noi{\bf II. Recursion and Differentiation Formulas}\sp

If we denote by $M$ multiplication by the independent variable, then the form
of the
kernel of $K$ shows that
\bq MK+KM=E\tn E,\label{Kcom}\eq
where in general we denote by $X\tn Y$ the operator with kernel $X(x)Y(y)$.
Applying
this twice we see that, with brackets denoting commutator as usual,
\[[M,\,K^2]=E\tn KE-KE\tn E.\]
It follows immediately
that if $Q_i:=(I-K^2)^{-1}E_i$ and $P_i:=(I-K^2)^{-1}KE_i$ then
\[[M,\,(I-K^2)^{-1}]=Q_0\tn P_0-P_0\tn Q_0.\]
Applying these operators to the function $E_j$ gives the recursion formula
\bq x\,Q_j(x)-Q_{j+1}(x)=Q_0(x)\,v_j-P_0(x)\,u_j,\label{Qrec}\eq
where we write $u_j$ for $u_{j,0}$ and $v_j$ for $v_{j,0}$. Taking inner
products
with $E_i$ gives
\bq u_{i+1,j}-u_{i,j+1}=u_i\,v_j-v_i\,u_j.\label{urec}\eq
To obtain the analogous relations for the $v_{i,j}$ we temporarily define
\[w_i:=((I-K^2)^{-1}KE,\,KE_i),\]
and take inner products with $KE_i$ in (\ref{Qrec}), obtaining
\[(MKE_i,\,Q_j)-v_{i,j+1}=v_i\,v_j-w_i\,u_j.\]
The identity $(I-K^2)^{-1}K^2=(I-K^2)^{-1}-I$ gives
\[w_i=u_i-(E,\,E_i),\]
and by (\ref{Kcom})
\[(MKE_i,\,Q_j)=-(KE_{i+1},\,Q_j)+(E,\,E_i)\,(E,\,Q_j)=
-v_{i+1,j}+(E,\,E_i)\,u_j.\]
Thus we obain
\bq v_{i+1,j}+v_{i,j+1}=u_i\,u_j-v_i\,v_j.\label{vrec}\eq

For the differentiation formulas we use the fact
\[{\pl\ov\pl t_k} E(x)\,E(y)=\hf\,(x^k+y^k)\,E(x)\,E(y)\]
and elementary algebra to deduce that for $k>0$
\bq{\pl K\ov\pl t_k}=\hf\,\sk E_i\tn E_j,\qquad
{\pl K\ov\pl t_{-k}}=\hf\,\smk E_i\tn E_j.\label{Kdiff}\eq
In the first sum we take $i,\,j\geq0$ and in the second $i,\,j\leq-1$. This
will be
our convention throughout. (Here we use the fact that $k$ is odd; the reader
will
find other such places later.) Since, with $t=t_k$ or $t_{-k}$,
\[{\pl\ph\ov\pl t}=\mbox{tr}\;(I+K)^{-1}{\pl K\ov\pl t}
+\mbox{tr}\;(I-K)^{-1}{\pl K\ov\pl t}=2\;\mbox{tr}\;(I-K^2)^{-1}{\pl K\ov\pl
t},\]
we find that
\bq{\pl\ph\ov\pl t_k}=\sk u_{i,j},\qquad{\pl\ph\ov\pl t_{-k}}=\smk u_{i,j}.
\label{phdiff}\eq
Notice especially the important fact
\bq{\pl\ph\ov\pl t_1}=u_0.\label{phdiff1}\eq

To obtain differentiation formulas for the $u_{i,j}$ and $v_{i,j}$ themselves
we
use
\[{\pl\ov\pl t_k}(I-K^2)^{-1}=(I-K^2)^{-1}{\pl K^2\ov\pl t_k}(I-K^2)^{-1}\]
and, by (\ref{Kdiff}),
\[{\pl K^2\ov\pl t_k}=K\,{\pl K\ov\pl t_k}+{\pl K\ov\pl t_k}\,K
=\hf\,\sk(K\,E_i\tn E_j+E_i\tn K\,E_j)\]
to deduce
\[{\pl\ov\pl t_k}(I-K^2)^{-1}=\hf\,\sk (P_i\tn Q_j+Q_i\tn P_j).\]
{}From this and the fact $\pl E_i/\pl t_k=E_{i+k}$ we deduce from the
definition
(\ref{uvdef}) that
\bq {\pl u_{p,q}\ov\pl t_k}=\hf\,\sk (u_{p,j}\,v_{q,i}+v_{p,j}\,u_{q,i})+
\hf\,(u_{p+k,q}+u_{p,q+k}).\label{udiff}\eq
If we introduce $R_i:=(I-K^2)^{-1}K^2E_i=Q_i-E_i$ then we find similarly first
\[{\pl\ov\pl t_k}(I-K^2)^{-1}K=\hf\,\sk(Q_i\tn R_j+P_i\tn P_j)+\hf\sk Q_i\tn
E_j\]
\[=\hf\,\sk(Q_i\tn Q_j+P_i\tn P_j)\]
and then
\bq {\pl v_{p,q}\ov\pl t_k}=\hf\,\sk (u_{p,j}\,u_{q,i}+v_{p,j}\,v_{q,i})+
\hf\,(v_{p+k,q}+v_{p,q+k}).\label{vdiff}\eq

In a completely analogous fashion, using the second part of (\ref{Kdiff}), we
obtain
formulas for differentiation with respect to the $t_{-k}$:
\bq {\pl u_{p,q}\ov\pl t_{-k}}=\hf\,\smk (u_{p,j}\,v_{q,i}+v_{p,j}\,u_{q,i})+
\hf\,(u_{p-k,q}+u_{p,q-k}),\label{udiff-}\eq
\bq {\pl v_{p,q}\ov\pl t_{-k}}=\hf\,\smk (u_{p,j}\,u_{q,i}+v_{p,j}\,v_{q,i})+
\hf\,(v_{p-k,q}+v_{p,q-k}).\label{vdiff-}\eq

\sp\noi{\bf III. The mKdV hierarchy}\sp

We begin by showing how to derive the first of the integrated mKdV equations,
\[{\pl\ph\ov \pl t_3}={\pl^3\ph\ov\pl t_1^3}-2\,({\pl\ph\ov\pl t_1})^3.\]
This will illustrate the procedure. By (\ref{phdiff1}) $\pl\ph/\pl t_1=u_0$,
and we differentiate twice more with respect to $t_1$, using (\ref{udiff}) and
(\ref{vdiff}). We find that the quantities $u_0,\,u_1,\,u_{1,1},\,v_0$ and
$v_1$
arise. But the recursion formulas (\ref{urec}) and (\ref{vrec}) allow us
to express two of these in terms of the others:
\[v_1=(u_0^2-v_0^2)/2,\quad u_{1,1}=u_2+u_0\,v_1-u_1\,v_0=u_2+\hf\,u_0\,
(u_0^2-v_0^2)-u_1\,v_0.\]
In the end the formula becomes
\[{\pl^3\ph\ov\pl t_1^3}={3\ov2}\,u_0^3+\hf\,u_0\,v_0^2+u_1\,v_0+u_2.\]
Now from (\ref{phdiff}), ${\pl\ph/\pl t_3}=2\,u_2-u_{1,1}$ and by the above
representation
of $u_{1,1}$ this may be written
\[{\pl\ph\ov\pl t_3}=-{1\ov2}u_0^3+{1\ov2}u_0\, v_0^2+u_1 \, v_0+u_2.\]
This gives
\[{\pl^3\ph\ov\pl t_1^3}-{\pl\ph\ov \pl t_3}=2\,u_0^3=2\,({\pl\ph\ov\pl
t_1})^3,\]
which is the desired equation.

The proof of the general formula (\ref{mKh}) follows from a series of three
lemmas.\sp

\noi{\it Lemma 1.} We have
\bq 2\,u_0\,{\pl u_0\ov\pl t_k}={\pl\ov\pl t_1}\sk u_i\,u_j.\label{Dinv}\eq
{\it Proof.} We begin by noting that from (\ref{udiff})
\[{\pl u_0\ov\pl t_k}=\sk u_i\,v_j+u_k\]
and, from (\ref{udiff}), (\ref{vdiff}), (\ref{urec}) and (\ref{vrec}),
\bq{\pl u_p\ov\pl t_1}=u_0\,v_p+u_{p+1},\qquad{\pl v_p\ov\pl t_1}=u_0\,u_p.
\label{t1der}\eq
We find that the right side of (\ref{Dinv}) equals
\[\sk [u_i\,(u_0\,v_j+u_{j+1})+u_j\,(u_0\,v_i+u_{i+1})]\]
\[=2\,u_0\,\sk u_i\,v_j+2\,\sk u_i\,u_{j+1}.\]
The last sum equals
\[u_0\,u_k-u_1\,u_{k-1}+u_2\,u_{k-2}-\cdots-u_{k-2}\,u_2+u_{k-1}\,u_1=
u_0\,u_k.\]
It follows that the right side of (\ref{Dinv}) equals the left side of
(\ref{Dinv}).
\sp

\noi{\it Lemma 2.} We have
\bq 2\,v_k=\sk(u_i\,u_j-v_i\,v_j).\label{vrep}\eq
{\it Proof.} By the recursion formulas (\ref{vrec}),
\[ v_{k,0}+v_{k-1,1}=u_0\,u_{k-1}-v_0\,v_{k-1}\]
\[-(v_{k-1,1}+v_{k-2,2})=-(u_1\,u_{k-2}-v_1\,v_{k-2})\]
\[\vdots\]
\[-(v_{2,k-2}+v_{1,k-1})=-(u_{k-2}\,u_1-v_{k-2}\,v_1)\]
\[v_{1,k-1}+v_{0,k}=u_{k-1}\,u_0-v_{k-1}\,v_0.\]
Adding gives (\ref{vrep}).\sp

\noi{\it Lemma 3.} We have for $k\geq1$
\bq {\pl\ph\ov\pl t_{k+2}}=D{\pl u_0\ov\pl t_k}-4\,u_0\,D^{-1}(u_0\,{\pl u_0\ov
\pl t_k}).\label{phirec}\eq
{\it Proof.} By Lemma 1 and the differentiation formula (\ref{udiff}) the
right side of (\ref{phirec}) equals
\[{\pl\ov\pl t_1}\big(\sk u_i\,v_j+u_k\big)-2\,u_0\,\sk u_i\,u_j,\]
and by (\ref{t1der}) this equals
\[\sk(u_i\,u_0\,u_j+u_0\,v_i\,v_j+u_{i+1}\,v_j)+u_0\,v_k+u_{k+1}-
2\,u_0\,\sk u_i\,u_j\]
\[=u_0\,\sk(v_i\,v_j-u_i\,u_j)+\sk u_{i+1}\,v_j+u_0\,v_k+u_{k+1}.\]
This is the right side of (\ref{phirec}). By (\ref{phdiff}) the left side
equals
$$u_{k+1}-(u_{1,k}-u_{2,k-1})-(u_{3,k-2}-u_{4,k-3})-
\cdots-(u_{k,1}-u_{k+1,0}),$$
and by (\ref{urec}) this equals
$$u_{k+1}+(u_1\,v_{k-1}-u_{k-1}\,v_1)+(u_3\,v_{k-3}-u_{k-3}\,v_3)+\cdots
+(u_k\,v_0-u_0\,v_k)$$
$$=u_{k+1}-\sum_{i+j=k}(-1)^i\,u_i\,v_j=u_{k+1}+\sk u_{i+1}\,v_j-u_0\,v_k.$$
Thus the difference between the right and left sides of (\ref{phirec}) equals
\[u_0\,\sk(v_i\,v_j-u_i\,u_j)+2\,u_0\,v_k\]
and by Lemma 2 this equals 0.\sp

The proof of (\ref{mKh}) is now immediate. In fact (\ref{phirec})
may be rewritten
\bq{\pl \ph\ov\pl t_{k+2}}=(D^2-4\,u_0\,D^{-1}\,u_0\,D)\,
{\pl\ph\ov\pl t_k},\label{phrec}\eq
and this together with (\ref{phdiff1}) gives (\ref{mKh}).\sp

\noi{\bf IV. The sinh-Gordon hierarchy}\sp

We begin by deriving (\ref{inv}).\sp

\noi{\it Lemma 4.} The operator $D^2-4\,u_0\,D^{-1}\,u_0\,D$ is invertible in
the
space of smooth functions all of whose derivatives are
rapidly decreasing as $t_1\ra-\iy$, and its inverse is given by (\ref{inv}).\sp

\noi{\it Remark.} The function $\ph$ and all the $u_{i,j}$ and $v_{i,j}$ belong
to the space of functions in the statement of the lemma.
\sp

\noi{\it Proof.} We have
\[D^2-4\,u_0\,D^{-1}\,u_0\,D=(I-4\,u_0\,D^{-1}\,u_0\,D^{-1})\,D^2.\]
Both factors on the right are invertible (the Neumann series inverts the first
factor) so the operator on the left is also, and its inverse is equal to
\[D^{-2}\,(I-4\,u_0\,D^{-1}u_0\,D^{-1})^{-1}
=\hf\,D^{-2}[(I-2\,u_0\,D^{-1})^{-1}+(I+2\,u_0\,D^{-1})^{-1}]\]
\[=\hf\,D^{-1}[(D-2\,u_0)^{-1}+(D+2\,u_0)^{-1}].\]
Since $(D+p)^{-1}=e^{-D^{-1}p}\,D^{-1}\,e^{D^{-1}p}$
and $D^{-1}\,u_0=\ph$, the above is equal to
\[\hf(D^{-1}\,e^{2\ph}\,D^{-1}\,e^{-2\ph}+
D^{-1}\,e^{-2\ph}\,D^{-1}\,e^{2\ph}).\]\sp

\noi{\it Lemma 5.} Relation (\ref{phrec}) holds for $k\leq-1$.\sp

\noi The proof of this is almost exactly the same as for $k\geq1$ and so is
omitted.\sp

Lemma 5 is equivalent to the statement that for $k=1,\,3,\,5,\cdots,$
\[{\pl \ph\ov\pl t_{-k+2}}=(D^2-4\,u_0\,D^{-1}\,u_0\,D)\,
{\pl\ph\ov\pl t_{-k}},\]
or by (\ref{inv}),
\[{\pl\ph\ov\pl t_{-k}}=\hf(D^{-1}\,e^{2\ph}\,D^{-1}\,e^{-2\ph}+D^{-1}\,
e^{-2\ph}\,D^{-1}\,e^{2\ph})\,{\pl \ph\ov\pl t_{-k+2}}.\]
This establishes (\ref{sgh}) by induction.

The case $n=1$ of (\ref{sgh}) is
\[{\pl\ph\ov\pl t_{-1}}=\hf(D^{-1}\,e^{2\ph}\,D^{-1}\,e^{-2\ph}+D^{-1}\,
e^{-2\ph}\,D^{-1}\,e^{2\ph})\,{\pl\ph\ov\pl t_1},\]
which gives (keep in mind that $D^{-1}$ is the antiderivative which vanishes at
$-\iy$)
\[4\,{\pl^2\ph\ov\pl t_{-1}\pl t_1}=4\,D\,{\pl\ph\ov\pl t_{-1}}=
2\,(e^{2\ph}\,D^{-1}\,e^{-2\ph}+e^{-2\ph}\,D^{-1}\,e^{2\ph})\,{\pl\ph\ov\pl
t_1}\]
\[=e^{2\ph}\,(1-e^{-2\ph})+e^{-2\ph}\,(e^{2\ph}-1)=2\,\sinh\,2\ph.\]
This is (\ref{sg}).

Finally we derive (\ref{I-K}). By (\ref{udiff-}) we have
\[{\pl^2\ph\ov\pl t_{-1}\pl t_1}={\pl u_0\ov\pl t_{-1}}=\um\,(1+\vm),\]
and so we know that
\[\um\,(1+\vm)=\hf\,\sinh\,2\ph.\]
Now we use a special case of (\ref{vrec}), $2\,\vm=\um^2-\vm^2,$
which has the more useful form
\[(1+\vm)^2=1+\um^2.\]
These equations can be solved for $\um$ and $\vm$, giving
\bq\um=\sinh\ph,\quad\vm=\cosh\ph-1.\label{uvm}\eq
Now we use the fact
$(I-K)^{-1}=(I-K^2)^{-1}+(I-K^2)^{-1}K$ and (\ref{Kdiff}) to obtain
$$-2{\pl\ov\pl t_1}\log\det\,(I-K)=((I-K)^{-1}E,\,E)=u_0+v_0.$$
Therefore by (\ref{udiff-}) and (\ref{vdiff-})
\[-2{\pl^2\ov\pl t_{-1}\pl t_1}\log\det\,(I-K)=\um(\vm+1+\um).\]
Using (\ref{uvm}) we find that the right side equals
$({e^{2\ph}-1)/2},$ which gives (\ref{I-K}).
\medskip
\begin{center}{\bf Acknowledgements}\end{center}
The authors wish to thank Albert Schwarz for bringing \cite{Z} to
their attention and Josef Dorfmeister and David Sattinger for helpful
comments on the sinh-Gordon hierarchy.
This work was supported in part by
the National Science Foundation through grants DMS--9303413 and DMS--9424292.

\end{document}